\documentclass[conference]{IEEEtran}
\usepackage{amsmath,amssymb,amsthm,mathrsfs,amsfonts,dsfont,stmaryrd,wasysym,marvosym}
\usepackage{mathtools}
\usepackage{algorithm}
\usepackage{algorithmic}
\usepackage{color}
\usepackage{graphicx}  
\usepackage{psfrag, xcolor}
\usepackage{enumitem}
\usepackage{subfigure,cite,epsfig,url}
\usepackage{booktabs}
\allowdisplaybreaks[4]
\newtheorem{theorem}{Theorem}
\newtheorem{lemma}{Lemma}

\newtheorem{definition}{Definition} 
\newtheorem{remark}{Remark} 
\newtheorem{example}{Example} 
\newcommand{\mc}{\mathcal} 
\newcommand{\mkv}{-\!\!\!\!\minuso\!\!\!\!-}

\graphicspath{{figs/}}

\newcommand{\W}{\mathsf{W}}

\newcommand{\mw}[1]{{\color{black}#1}}
\newcommand{\pe}[1]{{\color{black}#1}}
\newcommand{\az}[1]{{\color{black}#1}}

\DeclareMathOperator*{\argmin}{arg\,min}

\newcommand*{\coefaa}{\left({\bar{\sigma}_{XY_2}} {\bar{\sigma}_{Y_1Y_2}}  \! \! - \! {\bar{\sigma}_{XY_1}} {\bar \sigma_{Y_2}^2}\right)}
\newcommand*{\coefbb}{\left({\bar{\sigma}_{XY_1}} {\bar{\sigma}_{Y_1Y_2}}  \! \!  -\!   {\bar{\sigma}_{XY_2}} {\bar \sigma_{Y_1}^2}\right)}

\newcommand*{\coefa}{a}
\newcommand*{\coefb}{b}

\usepackage{tikz}
\usetikzlibrary{calc} 
\usetikzlibrary{patterns} 
\usetikzlibrary{arrows, decorations.markings}

\tikzstyle{vecArrow} = [thick, decoration={markings,mark=at position
	1 with {\arrow[semithick,color=black]{open triangle 60}}},
double distance=1.4pt, shorten >= 5.5pt,
preaction = {decorate},
postaction = {draw,line width=1.4pt, white,shorten >= 4.5pt}]
\tikzstyle{innerWhite} = [semithick, white,line width=1.4pt, shorten >= 4.5pt]

\tikzstyle{block}=[draw, rectangle, text centered, minimum width=2em, minimum height=3em]
\tikzstyle{sum}=[draw, circle, minimum size=.2cm]
\tikzstyle{rect}=[draw, dashed, rectangle, text centered, minimum width=12em, minimum height=8em]
\newcommand{\covmat}[1]{\mathbf{K}_{#1}}
\newcommand{\covmatbar}[1]{\bar{\mathbf{K}}_{#1}}
\newcommand{\crosscovmat}[2]{\mathbf{K}_{#1#2} }
\newcommand{\crosscovmatbar}[2]{\bar{\mathbf{K}}_{#1#2} }

\begin{document}
	\fontencoding{OT1}\fontsize{10}{11}\selectfont
\author{Pierre Escamilla\qquad Abdellatif Zaidi$^{\dagger}$ $^{\ddagger}$ \qquad Mich\`ele Wigger $^{\star}$  \vspace{0.3cm}\\
	$^{\dagger}$ Paris Research Center, Huawei Technologies, Boulogne-Billancourt, 92100, France\\
	$^{\ddagger}$ Universit\'e Paris-Est, Champs-sur-Marne, 77454, France\\
	$^{\star}$ LTCI, T\'el\'ecom Paris, Universit\'e Paris-Saclay, Palaiseau, 91120, France\\
	\{\tt pierre.escamilla@gmail.com, abdellatif.zaidi@u-pem.fr\}\\
	\{\tt michele.wigger@telecom-paristech.fr\}
}
\title{Some Results on the Vector Gaussian Hypothesis Testing Problem}
\maketitle
\begin{abstract}
This paper studies the problem of discriminating two multivariate Gaussian distributions \mw{in a distributed manner. Specifically, it characterizes in a special case the optimal type-II error exponent as a function of the available communication rate. As a side-result, the paper also presents  the optimal type-II error exponent of a slight generalization of the hypothesis testing against conditional independence problem where the marginal distributions under the two hypotheses can be different.}
\end{abstract}
\section{Introduction}\label{secI}
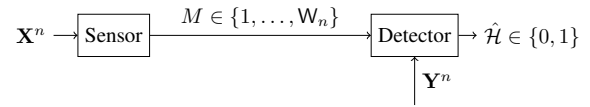
\begin{figure}[!ht]
	\begin{center}
		\scalebox{0.8}{
		\begin{tikzpicture}
		\node (in) at (0,0) [left]{$\mathbf X^{n}$};
		\node (enc) at (1,0) [block, minimum width=2em, minimum height=2em, align=center] {Sensor};	
		\node (det) at (6,0) [block, minimum width=2em, minimum height=2em, align=center] {Detector};
		\draw[->] (in) to (enc.west);
		\draw[->] (enc.east) to node[above] {$M \in \{1,\ldots, \W_n\}$} (det.west) ;
		\draw[->] ($(det.south)-(0,0.8)$) to node[right] {$\mathbf Y^{n}$} (det.south);
		\draw[->] (det.east) to ($(det.east)+(0.3,0)$) node[right] {$\hat{\mathcal{H}}\in\{0,1 \}$};	
		\end{tikzpicture}
		}
		\caption{Vector Gaussian hypothesis testing problem}
		\label{fig-system-model-P-1-vec}
	\end{center}
\end{figure}

Consider the single-sensor single-detector hypothesis testing scenario in Fig.~\ref{fig-system-model-P-1-vec}. The sensor observes a source sequence $\mathbf X^n\triangleq (\mathbf X_1,\ldots,\mathbf X_n)$  and communicates with the detector, who observes  source sequence $\mathbf Y^n \triangleq(\mathbf Y_{1},\ldots ,\mathbf Y_{n})$, over a noise-free bit-pipe of rate $R\geq 0$. Here, $n$ is a positive integer that  denotes the blocklength and the sequence of  pairs $\{(\mathbf X_{t}, \mathbf Y_{t})\}_{t=1}^n$ is independent and identically distributed (i.i.d) according to a jointly Gaussian distribution of zero-mean and of joint covariance matrix that depends on the  hypothesis {$\mc H \in \{0, 1\}$}. Under hypothesis
\begin{equation}\label{eq:H}
\mc H=0: \mw{\left\{\begin{pmatrix}\mathbf X_{t}\\ \mathbf Y_{t}\end{pmatrix} \right\}_{t=1}^n} \textnormal{  i.i.d. } \sim \mc N ( \mathbf 0 , \mathbf K ),
\end{equation}
and under hypothesis 
\begin{equation}\label{eq:H-bar}
\mc H=1: \mw{\left\{\begin{pmatrix}\mathbf X_{t}\\ \mathbf Y_{t}\end{pmatrix} \right\}_{t=1}^n} \textnormal{  i.i.d. } \sim \mc N ( \mathbf 0 , \bar{ \mathbf K } ).
\end{equation}
Based on its observations $\mathbf Y^n$ and the message it receives from the sensor, 
the Detector decides on the hypothesis by producing $\hat{\mc H} \in\{0,1\}$. The goal of  this decision is to  maximize the  exponential decrease  of the probability of type-II error (i.e., of guessing $\hat{\mc {H}} =0$   when $\mc H=1$), while ensuring that the probability of type-I error (i.e., guessing  $\hat{\mc {H}}=1$  when {$\mc H=0$}) goes to zero as $n \to \infty$ .

The described single-sensor single-detector problem has previously been studied in \cite{AC86, H87,SHA94,RW12} for various joint distributions on the i.i.d. observations. In particular,  \cite{RW12} identified \mw{the} largest type-II exponent that is achievable in a  setup that they termed \emph{testing against conditional independence}. \az{An explicit expression for the vector Gaussian case was recently found in~\cite{ZE19} (see Theorem 2 therein which actually provides the solution of a more general, distributed, setting)}. \mw{For all other  cases} a computable single-letter characterization of the largest achievable \mw{type-II error} exponent remains open. {This line of works has also been extended to multiple sensors \cite{H87,RW12,ZL14A,ZE19}, multiple detectors \cite{SWT18,EWZ19}, interactive terminals~\cite{TC08,XK12,KPD16A}, multi-hop networks~\cite{WT16,SWW17,EWZ18,ZL18,EZW19},  noisy channels~\cite{SW18A,SG17} and \mw{to} scenarios with privacy constraints~\cite{LSCV17,LSTC18,SGC18}.}

{In this paper we present a computable single-letter characterization of the  largest type-II error exponent achievable for the  Gaussian vector hypothesis testing problem  for a class of matrices $\mathbf{K}$ and $\bar{\mathbf{K}}$. Our converse proof starts from the known multi-letter  expression for this problem \cite{AC86}  and connects it to related results.
The achievability proof is based on the coding scheme proposed in \cite{SHA94}. }

We end this {introductory} section with some remarks on notation.
{When two random variables $(X,Y)$ are independent given a third random variable $Z$ (i.e. $P_{XYZ}= P_Z P_{X|Z}P_{Y|Z}$), \mw{we say $(X,Z,Y)$ form a Markov chain and  write $X \mkv Z \mkv Y$}.
Both $D(P_X\|P_{\bar{X}})$ and $D(X\|\bar{X})$ denote the Kullback-Leiber divergence between two pmfs $P_X$ and $P_{\bar{X}}$. 
 $h(\cdot)$, $I(\cdot;\cdot)$ and $I(\cdot;\cdot|\cdot)$ denote continuous entropy, mutual information and conditional mutual information. \mw{The set of all real numbers is denoted by $\mathbb{R}$.}
Boldface upper case letters denote random vectors or {deterministic} matrices, e.g., $\mathbf X$, where  {the} context should make the distinction clear. We denote the covariance \mw{matrix  of  a real-valued vector  $\mathbf X$ with distribution $P_{\mathbf X}$ by $\covmat{\mathbf X} = \mathbb E_{P_{\mathbf X}} [\mathbf X \mathbf X^\dagger ]$, where $^\dagger$ indicates the transpose operation}. Similarly, we denote the cross-correlation of two zero-mean vectors  $\mathbf X$ and $\mathbf Y$ with joint distribution $P_{\mathbf X \mathbf Y}$ by $\crosscovmat{\mathbf X}{\mathbf Y} = \mathbb E_{P_{\mathbf X \mathbf Y}} [\mathbf X \mathbf Y^\dagger ]$, the conditional covariance matrix of $\mathbf X$ given $\mathbf Y$ with p.d.f $ P_{\mathbf X \mathbf Y}$ and with p.d.f $\bar P_{\mathbf X \mathbf Y}$ by ${ \mathbf{K}}_{\mathbf X | \mathbf Y}  = \mathbb E_{ P_{\mathbf X \mathbf Y}} [\mathbf X \mathbf X^\dagger |\mathbf Y]$  and  $\bar{ \mathbf{K}}_{\mathbf X | \mathbf Y}  = \mathbb E_{\bar P_{\mathbf X \mathbf Y}} [\mathbf X \mathbf X^\dagger |\mathbf Y]$, respectively.
\pe{Finally, for a matrix $\mathbf M$,  we denote its inverse (if it exists) by $\mathbf{M}^{-1}$ its determinant (if it exists) by $|\mathbf{M}|$, its Moore-Penrose pseudo-inverse by $\mathbf M^+$ and its pseudo-determinant by $|\mathbf{M}|_+$.}
}

\section{ Formal problem statement }~\label{secII}
The sequences $\mathbf{X}^n$ and $\mathbf{Y}^n$ are as described before, where we denote by $m$ the dimension of each vector $\mathbf X_t$ and by $q$ the dimension of each vector $\mathbf Y_t$. 
The Sensor, which observes  $\mathbf X^n$ applies an encoding function
\begin{equation}\label{eq-phi-1-g}
\phi_{n}\colon  \mathbb R^{m\times n} \rightarrow \mc M =\{1,\ldots,\W_{n}\}
\end{equation} 
to this sequence and sends the resulting index 
\begin{equation}
M = \phi_{n}(\mathbf X^n)
\end{equation}
to the detector. 
Based on this message $M$ and its observation $\mathbf Y^n$, the detector then applies a decision function 
\begin{equation}\label{eq-psi-1-g}
\psi_{n} \colon \mc M \times \mathbb R^{q\times n} \rightarrow \{0,1\} 
\end{equation}
to decide on the hypothesis 
\begin{equation}\hat{\mc H} =\psi_{n} (M, \mathbf  Y^n ).
\end{equation} 
The Type-I and type-II error probabilities at the detector are defined as:
\begin{IEEEeqnarray}{rCl}
\alpha_{n} &\triangleq& \text{Pr}\big\{\hat{\mc H} = 1 \big|\mc H = 0 \}  \label{eq-type-I-error-constraint-at-Detector-P-1}\\
\beta_{n} &\triangleq&  \text{Pr}\big\{\hat{\mc H} = 0 \big |\mc H =1 \}. \label{eq-type-II-error-constraint-at-Detector-P-1}
\end{IEEEeqnarray}
\begin{definition}\label{def-achievability-P-2}
	Given rate $R\geq 0$, an error-exponent $\theta$ is said achievable if for  all  blocklengths $n$ there exist functions $\phi_{n}$ and $\psi_{n}$ as in \eqref{eq-phi-1-g} and \eqref{eq-psi-1-g} so that the following limits hold:
	\begin{subequations}\label{eq-achievability-P-1}
		\begin{equation}
		\lim_{n\to \infty} \alpha_{n} = 0,
		\label{eq-definition-constant-constraints-typeI-error-p2p-p1}
		\end{equation}
		\begin{equation}
		\theta  \leq   \varliminf_{n \to \infty}-\frac{1}{n} \log \beta_{n}
		\end{equation}
		and 
		\begin{equation}\label{eq-rate-constraint-p1}
		\varlimsup_{n\to \infty} \frac{1}{n}\log_2 \mathsf{W}_{n} \leq R .
		\end{equation}
	\end{subequations}
\end{definition}

\begin{definition}[Exponent-rate function] \label{def-rate-error-exponent-function} For any rate $R\geq 0$, the \emph{exponent-rate function} \mw{$E(R)$} is the supremum of the set of all achievable error-exponents.
\end{definition}
In essence, the problem of vector Gaussian hypothesis testing that we study here amounts to discriminating two covariance matrices. As we already mentioned the solution of this problem is known only in few special cases, namely the cases of testing against independence and testing against conditional independence \mw{\cite{AC86}, \cite{RW12}, and \az{\cite[Theorem 2]{ZE19}}.} 

\section{Optimal exponent for a class of \az{covariance matrices}}
Let $\covmat{\mathbf X}$ and $\covmatbar{\mathbf X}$ be $m$-by-$m$ dimensional matrices, $\covmat{\mathbf Y}$ and $\covmatbar{\mathbf Y}$ be $q$-by-$q$ dimensional matrices, and $\crosscovmat{\mathbf X }{\mathbf  Y}$ and $\crosscovmatbar{\mathbf X }{\mathbf  Y}$ be $m$-by-$q$ dimensional matrices such that 
\begin{equation}
\mathbf{K}  = \left [ \begin{matrix} \covmat{\mathbf X}  & \crosscovmat{\mathbf X }{\mathbf  Y}\\ \crosscovmat{\mathbf X }{\mathbf  Y}^\dagger & \covmat{\mathbf Y } \end{matrix}\right ]
\textnormal{and }
\mathbf{\bar K}  =  \left [ \begin{matrix} \covmatbar{\mathbf X}  & \crosscovmatbar{\mathbf X }{\mathbf Y}\\ \crosscovmatbar{\mathbf X }{\mathbf  Y}^\dagger & \covmatbar{\mathbf Y }\end{matrix}\right ].
\end{equation}
Further, define the condition   $\textnormal{C}$: 
\begin{IEEEeqnarray} {rl}
\textnormal{C}\colon \crosscovmat{\mathbf X }{\mathbf  Y}\! &=\! \argmin_{\begin{subarray}{c} \mathbf {G}   \end{subarray} }  \log {\left|\!  \left[ \begin{matrix}
	\mathbf{I} &\mathbf 0\\ \mathbf 0 &  \crosscovmatbar{\mathbf X }{\mathbf  Y} \bar{\mathbf K}_{\mathbf Y}^{-1}  \end{matrix}\right] \!\! \bar{\mathbf K}\!\!\left[ \begin{matrix}
	\mathbf{I} &\mathbf 0\\ \mathbf 0 &  \crosscovmatbar{\mathbf X }{\mathbf  Y} \bar{\mathbf K}_{\mathbf Y}^{-1}  \end{matrix}\right]^\dagger \!\right |_{\pe +} } \nonumber \\ &\quad - \log{\left |  \mathbf \Gamma \right | } \nonumber \\ &  +  \textnormal{Tr}\left \{ \left ( \left[ \begin{matrix}
	\mathbf{I} &\mathbf 0\\ \mathbf 0 &  \crosscovmatbar{\mathbf X }{\mathbf  Y} \bar{\mathbf K}_{\mathbf Y}^{-1}  \end{matrix}\right] \bar{\mathbf K} \left[ \begin{matrix}
	\mathbf{I} &\mathbf 0\\ \mathbf 0 &  \crosscovmatbar{\mathbf X }{\mathbf  Y} \bar{\mathbf K}_{\mathbf Y}^{-1}  \end{matrix}\right]^\dagger \right )^{\pe +} \!\!\!\!\!{  \mathbf \Gamma }  \right \} \!\! \nonumber \\* \label{eq-condition-1}
\end{IEEEeqnarray} 
where the minimum is over all $m$-by-$q$ matrices $\mathbf G$ such that the matrix 
 \begin{equation}
 \mathbf \Gamma  \triangleq  \left[ \begin{matrix}
 \mathbf K_{\mathbf X} & \mathbf G^\dagger \bar{\mathbf K}_{\mathbf Y}^{-1} \crosscovmatbar{\mathbf X }{\mathbf  Y}^\dagger\\ \crosscovmatbar{\mathbf X }{\mathbf  Y} \bar{\mathbf K}_{\mathbf Y}^{-1} \mathbf G &  \crosscovmatbar{\mathbf X }{\mathbf  Y} \bar{\mathbf K}_{\mathbf Y}^{-1}\mathbf K_{\mathbf Y} \bar{\mathbf K}_{\mathbf Y}^{-1} \crosscovmatbar{\mathbf X }{\mathbf  Y}^\dagger  \end{matrix}\right] .
 \end{equation}
 is positive semi-definite, \az{i.e.,}
 \begin{equation}
\mathbf{ \Gamma } \succeq 0. 
 \end{equation}
The following theorem provides an explicit analytic expression of the exponent-rate function of the vector Gaussian hypothesis testing problem of Figure~\ref{fig-system-model-P-1-vec} when  condition $\text{C}$ in \eqref{eq-condition-1} is \az{fulfilled.}
\begin{theorem}\label{theorem-single-letter} If $\textnormal{C}$ is satisfied, then
	\begin{IEEEeqnarray}{rl}\label{eq-vectorial-rate-exponent-function}
	E&(R) \nonumber \\ &= \frac{m}{2} +\frac{q}{2} + \frac{1}{2} \log \frac{|\covmatbar{\mathbf Y}|}{|\covmat{\mathbf Y}|}  +  \frac{1}{2} \textnormal{Tr}  \left ( \covmatbar{\mathbf Y}^{-1} \covmat{\mathbf Y}\right ) \nonumber \\
	&  + \frac{1}{2} \log  | \covmatbar{\mathbf X | \mathbf Y}| - \log |\covmat{\mathbf X} - \crosscovmat{\mathbf X}{\mathbf Y}  \covmatbar{\mathbf Y}^{-1} \nonumber \\ 
	& \quad  \times \crosscovmatbar{\mathbf X}{\mathbf Y} ^\dagger  (  \crosscovmatbar{\mathbf X}{\mathbf Y} \covmatbar{\mathbf Y}^{-1} \covmat{\mathbf Y} \covmatbar{\mathbf Y}^{-1} \crosscovmatbar{\mathbf X}{\mathbf Y}^\dagger  )^{\pe +}\crosscovmatbar{\mathbf X}{\mathbf Y} \covmatbar{\mathbf Y}^{-1}  \crosscovmat{\mathbf X}{\mathbf Y} ^\dagger | \nonumber \\ 
	& +  \frac{1}{2} \textnormal{Tr} \left ( \covmatbar{\mathbf X | \mathbf Y}^{-1}  \times \right. \nonumber \\
 	&  \left (\covmat{\mathbf X} - \crosscovmat{\mathbf X}{\mathbf Y}  \covmatbar{\mathbf Y}^{-1}  \crosscovmatbar{\mathbf X}{\mathbf Y} ^\dagger  \left (  \crosscovmatbar{\mathbf X}{\mathbf Y} \covmatbar{\mathbf Y}^{-1} \covmat{\mathbf Y} \covmatbar{\mathbf Y}^{-1} \crosscovmatbar{\mathbf X}{\mathbf Y}^\dagger  \right )^{\pe +} \right . \nonumber \\
	&\left .\left . \times \crosscovmatbar{\mathbf X}{\mathbf Y} \covmatbar{\mathbf Y}^{-1}  \crosscovmat{\mathbf X}{\mathbf Y} ^\dagger \right  ) \right )\nonumber \\   
	& +  \max  \min \left \{ R + \frac{1}{2} \log  \big | \mathbf{I} - \mathbf \Omega \mathbf K_{\mathbf X | \mathbf Y} \big |  , \right.\nonumber \\  
	&   \frac{1}{2}\log \Big |  \mathbf{I}  +  \mathbf  \Omega  \crosscovmat{\mathbf X}{\mathbf Y} \nonumber \\ &\left. \qquad \times \Bigg (  \covmat{\mathbf Y}^{-1} -  \covmat{\mathbf Y}^{-1} \covmatbar{\mathbf Y}    \crosscovmatbar{\mathbf X}{\mathbf Y} ^{+} \crosscovmatbar{\mathbf X}{\mathbf Y} \covmatbar{\mathbf Y}^{-1}  \Bigg)   \crosscovmat{\mathbf X}{\mathbf Y} ^\dagger \Big |  \right \},\label{eq:result} 
	\end{IEEEeqnarray}
	where the maximization in the last term is over all matrices $\mathbf 0 \preceq \mathbf  \Omega \preceq  \mathbf K_{\mathbf X | \mathbf Y}^{-1}$ and where $\bar{\mathbf K}_{\mathbf X \mathbf Y}^+$ designates the Moore-Penrose pseudo inverse of $\bar{\mathbf K}_{\mathbf X \mathbf Y}$. \end{theorem}	
\begin{IEEEproof}
	See Section~\ref{sec-proof-of-theorem}.
\end{IEEEproof}
\begin{remark}
The theorem recovers the result \az{of} \cite[Theorem 7]{RW12}  in the special case of \mw{testing} against independent and  $m=q=1$. In this case, when the \az{distribution} $P_{XY}$ used under \az{the null hypothesis} $\mathcal{H}=0$ describes the channel   $Y=X+N$ with $X$ and $N$ independent \az{Gaussian} \az{both} with zero mean and \az{respective} variances $\sigma_X^2$ and  $\sigma_N^2$, and  the joint law $\bar{P}_{XY}$ under $\mathcal{H}=1$ describes a pair of independent Gaussians of variances \mw{$\sigma_X^2+\sigma_N^2$} and $\sigma_N^2$, then:
		\begin{equation}
	E(R) = \frac{1}{2} \log \left( \frac{\sigma^2_X + \sigma^2_N}{\sigma^2_N+2^{-2R}\sigma^2_X} \right).
	\end{equation}
	\end{remark} 

\section{Proof of Theorem~\ref{theorem-single-letter}}\label{sec-proof-of-theorem}

We first derive an auxiliary result. Consider a slight generalization of the discrete memoryless single-sensor single-detector hypothesis testing against conditional independence problem where the  marginals are not identical under the two hypotheses. Specifically, consider 
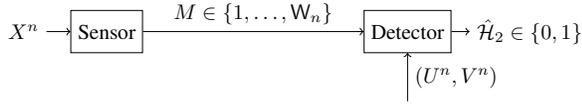
\begin{figure}[!ht]
	\begin{center}
		\scalebox{0.8}{
		\begin{tikzpicture}
		\node (in) at (0,0) [left]{$ X^{n}$};
		\node (enc) at (1,0) [block, minimum width=2em, minimum height=2em, align=center] { Sensor};	
		\node (det) at (6,0) [block, minimum width=2em, minimum height=2em, align=center] {Detector};
		\draw[->] (in) to (enc.west);
		\draw[->] (enc.east) to node[above] {$M\in \{1,\ldots,\W_n\}$} (det.west) ;
		\draw[->] ($(det.south)-(0,0.8)$) to node[right] {$ (U^n,V^n)$} (det.south);
		\draw[->] (det.east) to ($(det.east)+(0.3,0)$) node[right] {$\hat{\mathcal{H}}_2\in\{0,1 \}$};	
		\end{tikzpicture}
		}
		\caption{Hypothesis testing with two detector observations.}
		\label{fig-system-model-P-2}
	\end{center}
\end{figure}
 the problem of Figure~\ref{fig-system-model-P-2} where under  
 \begin{subequations}\label{eq:TACI}
 \begin{IEEEeqnarray}{rCl}
\mc H &=&0:\quad\!\!\{(X_t,U_t,V_t)\}_{t=1}^n \textnormal{ i.i.d. } \sim P_{ X U V} \\
\mc H &=&1:\quad\!\! \{(X_t,U_t,V_t)\}_{t=1}^n \textnormal{ i.i.d. } \sim  \bar P_{ X U  V} \!= \! \bar P_{ U} \bar P_{ X | U} \bar P_{ V | U}. \nonumber \\
\end{IEEEeqnarray}
\end{subequations}
\mw{for  arbitrary  distributions $P_{XUV}$, $\bar{P}_U$, $\bar{P}_{X|U}$, and $\bar{P}_{V|U}$. }

In this new setup the message $M$ and the decision $\hat{\mathcal H}$ are obtained as described in the previous section if the observation $\mathbf Y^n$ is replaced by the pair of sequences $\mathbf{U}^n\triangleq (\mathbf{U}_1,\ldots, \mathbf{U}_n)$ and $\mathbf{V}^n\triangleq (\mathbf{V}_1,\ldots, \mathbf{V}_n)$. Type-I and type-II error probabilities, achievable exponents, and exponent-rate function are defined as in Section~\ref{secII}.
\begin{lemma}
	\label{lemma-1}
	 If the joint \mw{distribution} $P_{XU}$  under the null  hypothesis satisfies
	\begin{equation}\label{eq-zero-rate_is_centralized}
	P_{XU} =  \argmin_{\begin{subarray}{c}
		\tilde P_{XU}: \tilde P_X=P_X \\ \tilde P_U=P_U  
		\end{subarray}}  D(\tilde P_{XU} \| \bar P_{XU}),
		\end{equation}  the rate exponent function is given by
	\begin{IEEEeqnarray}{rCl}\label{eq-general-form}
E(R) &=&	D(P_{XU}\| \bar P_{XU}) + \mathbb{E}_{P_U}\big[D(P_{V|U}\|\bar P_{V|U}) \big]\nonumber \\
&&+ \max I(S;V|U)
	\end{IEEEeqnarray} where in \eqref{eq-general-form} the maximization is over all conditionals p.m.f.s $P_{S|X}$ for which $I(S;X|U) \leq R$.
\end{lemma}
\begin{IEEEproof}[Proof of Lemma~\ref{lemma-1}]
	By {\cite[Theorem~4]{AC86}}:
\begin{equation}\label{step-0}
E(R) = \varliminf_{n \to \infty}  E_{n}(R),
\end{equation}
where \begin{equation}
E_{n}(R) \triangleq \max_{\begin{subarray}{c} \phi_n\;:\\ \log_2 |\phi_n| \leq nR \end{subarray}}  \frac{1}{n}D(P_{\phi_n(X^n)U^nV^n}\|\bar P_{\phi_n( X^n) U^n V^n}).
\end{equation}
\az{Next, notice} that by the chain rule for KL divergence, the data processing inequality,  and some simple manipulations, \az{we have}
\begin{IEEEeqnarray}{rCl}
\lefteqn{D(P_{\phi_n(X^n)U^nV^n}\|\bar P_{\phi_n( X^n) U^n V^n}) }  \nonumber \\
& = & D(P_{\phi_n(X^n)U^n }\| \bar{P}_{\phi_n(X^n)U^n } ) \nonumber\\
&&  + \mathbb{E}_{P_{\phi_n(X^n)U^n}}\big[ D(P_{V^n|\phi_n(X^n)U^n} \| \bar{P}_{V^n|U^n}) \big]\\
& = & D(P_{\phi_n(X^n)U^n }\| \bar{P}_{\phi_n(X^n)U^n } ) + I(V^n; \phi_n(X^n)|U^n)\nonumber\\
&&   + n \mathbb{E}_{P_{U}}\big[D(P_{V|U }\| \bar{P}_{V|U } ) \big]\\
& \stackrel{(a)}{\leq} & nD(P_{XU }\| \bar{P}_{XU } ) + I(V^n; \phi_n(X^n)|U^n)\nonumber\\
&&   + n \mathbb{E}_{P_{U}}\big[D(P_{V|U }\| \bar{P}_{V|U } ) \big], \label{eq:KL_equivalence}
	\end{IEEEeqnarray}
	where $(a)$ holds by the data-processing inequality for KL-divergence and because $X^n$ \mw{and} $U^n$ are i.i.d. 
	
We can thus bound $E(R)$ as:	
\begin{IEEEeqnarray}{rCl}\label{step-1}
	E(R)  & \leq & D(P_{XU }\| \bar{P}_{XU } )  + \mathbb{E}_{P_{U}}\big[D(P_{V|U }\| \bar{P}_{V|U } ) \big] \nonumber \\
	& & + \varliminf_{n\to \infty} \max_{\begin{subarray}{c} \phi_n\;:\\ \log_2 |\phi_n| \leq nR \end{subarray}} \frac{1}{n} I(V^n; \phi_n(X^n)|U^n).
	\end{IEEEeqnarray}
	Next we use that  by {\cite[Theorem 4]{AC86}} and \cite[Theorem 3]{RW12}  both sides of 
	\begin{IEEEeqnarray}{rCl}\label{eq:ISUV}
	\lefteqn{\varliminf_{n \to \infty}\!\! \max_{\begin{subarray}{c} \phi_n\;:\\ \log_2 |\phi_n| \leq nR \end{subarray}} \!\!\!\! \frac{1}{n}\cdot(I(\phi_{n}(X^n);V^n|U^n))} \nonumber\\
	&= & 	\max_{\begin{subarray}{c} P_{S|X}\;:\\ I(S;X|U) \leq R\end{subarray} } I(S;V|U)\hspace{2cm}
	\end{IEEEeqnarray} characterize the optimal type-II error exponent of a hypothesis testing against conditional independence problem at rate $R$, and thus coincide.

%
 
Combining \eqref{step-0} and \eqref{eq:ISUV} we obtain:
	\begin{IEEEeqnarray}{rCl}
		E(R) &\leq &	D(P_{XU}\| \bar P_{XU}) + \mathbb{E}_{P_U}\big[D(P_{V|U}\|\bar P_{V|U}) \big]\nonumber \\
		&+&\max I(S;V|U).
	\end{IEEEeqnarray}
The reverse inequality follows from the achievable type-II error exponent of Shimokawa-Han-Amari (SHA) \cite{SHA94} \mw{(see \cite[Section~IV]{SWW19} for an analysis)} which states that for every choice of the conditional distribution $P_{S|X}$ satisfying $R \geq I(S;X|U,V)$ the following lower bound holds:
	\begin{align}
	E(R) &\geq  \min \Bigg\{ \min_{\substack{\tilde{P}_{SXY}:\\\tilde{P}_{SX}=P_{SX}\\\tilde{P}_{SUV}=P_{SUV}}}\hspace{-0.2cm} D(\tilde{P}_{SXUV}\|P_{S|X}\bar P_{XU}\bar P_{V|U}), \nonumber \\
	& \hspace{0cm} \min_{\substack{\tilde{P}_{SXUV}:\\\tilde{P}_{SX}=P_{SX}\\\tilde{P}_{UV}=P_{UV}\\H(S|UV) \leq H_{\tilde{P}_{SUV}}(S|UV)}}\hspace{-0.9cm} D(\tilde{P}_{SXUV}\|P_{S|X}\bar P_{XU}\bar P_{V|U})\nonumber\\[-9ex] & \hspace{4.5cm} + R-I(S;X|UV)\Bigg\}\label{ach}
	\end{align}	
	\mw{where the mutual information $I(S;X|UV)$ is calculated according to $P_{S|X}P_{UVX}$.}
 \mw{In what follows we show that the SHA result implies that the error exponent on the RHS of \eqref{eq-general-form} is achievable.} 
\mw{As  in \cite{RW12}, we restrict to distributions $P_{S|X}$  satisfying}
 \begin{equation}\label{eq:rate-diff}
 R \geq I(S;X|U) 
 \end{equation}
 and drop the condition $H(S|U,V) \leq H_{\tilde{P}_{SUV}}(S|U,V)$ in the minimization. These changes can lead to a smaller exponent \mw{than  in \cite{SHA94},} and thus the resulting exponent is still achievable. 
 
By the Markov chain {$S \mkv X \mkv (U,V)$}, Condition \eqref{eq:rate-diff} implies 
 \begin{equation}\label{eq:s2}
 R- I(S;X|U,V)  \geq I(S;V|U).
 \end{equation} 
   Moreover, by the chain rule and the nonnegativity and  convexity of KL divergence, for any $\tilde{P}_{SXUV}$:
   \begin{IEEEeqnarray}{rCl} 
\lefteqn{   	D(\tilde{P}_{SXUV}\|P_{S|X}\bar{P}_{XU}\bar{P}_{V|U}) } \ \nonumber \\
& \geq & 	D(\tilde{P}_{XUV}\|\bar{P}_{XU}\bar{P}_{V|U}) \\
&=  &  D(\tilde{P}_{XU}\|\bar  P_{XU}) + \mathbb{E}_{\tilde{P}_{XU}}[ D( \tilde{P}_{V|XU} \| \bar{P}_{V|U})] \\
&\geq   &  D(\tilde{P}_{XU}\|\bar  P_{XU}) + \mathbb{E}_{\tilde{P}_{U}}[ D( \tilde{P}_{V|U} \| \bar{P}_{V|U})] .\label{eq:s1}
	   \end{IEEEeqnarray}
	   By \eqref{eq:s2} and \eqref{eq:s1} and  since the   \mw{second} minimization in \eqref{ach} is over distributions $\tilde{P}_{SXUV}$  satisfying $\tilde{P}_{UV}=P_{UV}$, we conclude that under conditions \eqref{eq-zero-rate_is_centralized} and \eqref{eq:rate-diff} the second term in \eqref{ach} is    lower bounded by
	   \begin{equation}
	\theta\triangleq    D({P}_{XU}\|\bar P_{XU}) + \mathbb{E}_{{P}_{U}}[ D( {P}_{V|U} \| \bar{P}| _{V|U})] +  I(S;V|U).
	   \end{equation}
We now  lower bound the first term \mw{in \eqref{ach}.} By the chain rule and the nonnegativity and  convexity of KL divergence, for any $\tilde{P}_{SXUV}$ where $\tilde{P}_{SX}=P_{S|X} P_X$:
   \begin{IEEEeqnarray}{rCl} 
   	\lefteqn{   	D(\tilde{P}_{SXUV}\|P_{S|X}\bar{P}_{XU}\bar{P}_{V|U}) } \ \nonumber \\
   	&=  &  D(\tilde{P}_{XU}\|\bar P_{XU}) + \mathbb{E}_{\tilde{P}_{XU}}[ D( \tilde{P}_{SV|XU} \| {P}_{S|X}\bar{P}_{V|U})]  \nonumber  \\ 
   	 	&\geq   &  D(\tilde{P}_{XU}\|\bar P_{XU}) + \mathbb{E}_{\tilde{P}_{U}}[ D( \tilde{P}_{SV|U} \| {P}_S \bar{P}_{V|U})]. \label{eq:t1}\IEEEeqnarraynumspace
   \end{IEEEeqnarray}
   
Where in the last inequality we used that  $\sum_{x}\! \tilde{P}_X(x) P_{S|X}(s|x) \! \! \! =\! \! \! \mw{{P}}_{S}(s)$ because $\tilde{P}_X\! \! \!=\! \! \!P_X$.
We now notice that the \mw{first} miniminization in \eqref{ach} is only over distributions  $\tilde{P}_{SXUV}$ satisfying $\tilde{P}_{SUV}=P_{SUV}$ and therefore: 
   \begin{IEEEeqnarray}{rCl} 
\lefteqn{\mathbb{E}_{\tilde{P}_{U}}[ D( \tilde{P}_{SV|U} \| {P}_{S}\bar{P}_{V|U})] } \qquad \nonumber \\
&= & \mathbb{E}_{{P}_{U}}[ D( {P}_{SV|U} \| {P}_{S}\bar{P}_{V|U})] \\
& =& I(S;V|U)+ \mathbb{E}_{{P}_{U}}[ D( {P}_{V|U} \| \bar{P}_{V|U})] .\label{eq:t2}
   \end{IEEEeqnarray}
Combining  \eqref{eq:t1} and \eqref{eq:t2}, we conclude that under Condition \mw{\eqref{eq-zero-rate_is_centralized}} also  the first term in the minimization in \eqref{ach} is  lower bounded by $\theta$. This establishes the achievability of the right-hand side of \eqref{eq-general-form}. 
 \end{IEEEproof}
 \medskip
 
We  turn to the proof of Theorem~\ref{theorem-single-letter}.
Define \begin{IEEEeqnarray}{rCl}
\mathbf U &=&\mathbb E_{\bar P}[\mathbf X|\mathbf Y] \\
\mathbf V &=& \mathbf Y
\end{IEEEeqnarray} and notice that under  $\mc H=1$ they satisfy the Markov chain\begin{equation}\label{eq:markov0}
\mathbf X \mkv \mathbf U \mkv \mathbf V.
\end{equation}  
In what remains, we assume that  \mw{instead of  $\mathbf{Y}^n$ the decoder observes the pair of sequences $(\mathbf{U}^n, \mathbf{V}^n)$} which is i.i.d according to the joint distribution of $(\mathbf{U},\mathbf{V})$. This new system  is depicted in Figure~\ref{fig-system-model-P-2}. Since there is a bijection between $\mathbf{Y}^n$  and  \mw{$(\mathbf{U}^n, \mathbf{V}^n)$, the error exponent of the new system coincides with the error exponent of the original system}. Moreover, by the Markov chain \eqref{eq:markov0} the new system is a generalized testing against conditional independen\mw{ce} problem as described in \eqref{eq:TACI}. We next argue that under condition C in Theorem~\ref{theorem-single-letter} and because $P_{\mathbf{X}^n\mathbf{U}^n \mathbf{V}^n}$ and $\bar P_{\mathbf{X}^n\mathbf{U}^n \mathbf{V}^n}$ are multivariate Gaussian distributions,  the new system also satisfies Condition~\eqref{eq-zero-rate_is_centralized} in Lemma~\ref{lemma-1}. The optimal exponent  $E(R)$ will then follow immediately from this Lemma~\ref{lemma-1}. 

\noindent To show that for  multivariate Gaussian distributions $P_{\mathbf{X}^n\mathbf{U}^n \mathbf{V}^n}$ and $\bar P_{\mathbf{X}^n\mathbf{U}^n \mathbf{V}^n}$  condition C  in  \eqref{eq-condition-1} implies \eqref{eq-zero-rate_is_centralized},  we first show that under this Gaussian assumption the minimizer of \begin{equation}
\argmin_{\begin{subarray}{c}
	\tilde P_{\mathbf X\mathbf U}: \tilde P_{\mathbf X}=P_{\mathbf X}  \\ \tilde P_{\mathbf U} = P_{\mathbf U}   
	\end{subarray}}  D(\tilde P_{\mathbf X\mathbf U} \| \bar P_{\mathbf X\mathbf U})
\end{equation}
is a multivariate Gaussian distribution. \mw{To see this fix any distribution $\tilde P_{\mathbf X\mathbf U}$ with $\tilde P_{\mathbf X} =P_{\mathbf X}$ and $\tilde P_{\mathbf U} = P_{\mathbf U}$ and let  $\tilde P_{\mathbf X\mathbf U}^G$  be a multivariate Gaussian distribution  with  same covariance matrix as $\tilde P_{\mathbf X\mathbf U}$. Then:
\begin{IEEEeqnarray}{rCl}
	D(\tilde P_{\mathbf X\mathbf U} \| \bar P_{\mathbf X\mathbf U}) &=& -h(\tilde P_{\mathbf X\mathbf U})  -  \mathbb E_{\tilde P} \left [\log \bar P_{\mathbf X\mathbf U} \right ] \nonumber \\ 
		&\geq& -h( \tilde P_{\mathbf X\mathbf U}^G)  -  \mathbb E_{\tilde P} \left [ \log \bar P_{\mathbf X\mathbf U} \right ] \nonumber\\
	&=& -h( \tilde P_{\mathbf X\mathbf U}^G)  -  \mathbb E_{\tilde P^G} \left [  \log \bar P_{\mathbf X\mathbf U} \right],
\end{IEEEeqnarray}  where the inequality holds because a Gaussian distribution maximizes differential entropy under a fixed covariance matrix constraint and where the last equality holds because $\mathbb{E}[\log \bar P_{UX}]$ only depends on the covariance matrix of $(U,X)$ which is the same under $\tilde{P}$ and $\tilde{P}^G$.
By straightforward algebra, it can then be shown that if condition C in \eqref{eq-condition-1} holds, then $P_{\mathbf{UX}}$ is the multivariate Gaussian distribution that minimizes \eqref{eq-zero-rate_is_centralized}. }

We conclude that the optimal exponent $E(R)$ is given by \eqref{eq-general-form} in Lemma~\ref{lemma-1}. We evaluate \eqref{eq-general-form} for our problem. For simplicity, we rewrite 
\begin{IEEEeqnarray}{rCl}
\lefteqn{D(P_{\mathbf X \mathbf U}\| \bar P_{\mathbf X\mathbf U}) + \mathbb{E}_{P_{\mathbf U}}\big[D(P_{\mathbf V|\mathbf U}\|\bar P_{\mathbf V| \mathbf U}) }\quad \nonumber\\
& = & D(P_{\mathbf{U}\mathbf{V}} || \bar{P}_{\mathbf{U}\mathbf{V}})   +  \mathbb{E}_{P_{\mathbf U}}\big[ D( P_{{\mathbf X|\mathbf U}}|| \bar{P}_{{ \mathbf X| \mathbf U}} ) \big], \label{eq:rewrite}
\end{IEEEeqnarray} 
 and proceed to compute 
\begin{equation}\label{eq-const-term-1}
D(P_{\mathbf{U}\mathbf{V}} || \bar{P}_{\mathbf{U}\mathbf{V}}) = \frac{q}{2} + \frac{1}{2} \log \frac{|\covmatbar{\mathbf Y}|}{|\covmat{\mathbf Y}|}  +  \frac{1}{2} \textnormal{Tr}  \left ( \covmatbar{\mathbf Y}^{-1} \covmat{\mathbf Y}\right )
\end{equation} 
and
\begin{IEEEeqnarray}{rl}
 D(& P_{\mathbf{X}}|| \bar{P}_{\mathbf{X}} | \mathbf{U} )\!=\! \frac{m}{2}  + \frac{1}{2} \log  | \covmatbar{\mathbf X | \mathbf Y}| - \log |\covmat{\mathbf X} - \crosscovmat{\mathbf X}{\mathbf Y}  \covmatbar{\mathbf Y}^{-1} \nonumber \\ 
 & \quad  \times \crosscovmatbar{\mathbf X}{\mathbf Y} ^\dagger  (  \crosscovmatbar{\mathbf X}{\mathbf Y} \covmatbar{\mathbf Y}^{-1} \covmat{\mathbf Y} \covmatbar{\mathbf Y}^{-1} \crosscovmatbar{\mathbf X}{\mathbf Y}^\dagger  )^{\pe +}\crosscovmatbar{\mathbf X}{\mathbf Y} \covmatbar{\mathbf Y}^{-1}  \crosscovmat{\mathbf X}{\mathbf Y} ^\dagger | \nonumber \\ 
 & +  \frac{1}{2} \textnormal{Tr} \left ( \covmatbar{\mathbf X | \mathbf Y}^{-1}  \times \right. \nonumber \\
 &  \left (\covmat{\mathbf X} - \crosscovmat{\mathbf X}{\mathbf Y}  \covmatbar{\mathbf Y}^{-1}  \crosscovmatbar{\mathbf X}{\mathbf Y} ^\dagger  \left (  \crosscovmatbar{\mathbf X}{\mathbf Y} \covmatbar{\mathbf Y}^{-1} \covmat{\mathbf Y} \covmatbar{\mathbf Y}^{-1} \crosscovmatbar{\mathbf X}{\mathbf Y}^\dagger  \right )^{\pe +} \right .\nonumber \\
 &\left . \left . \times \crosscovmatbar{\mathbf X}{\mathbf Y} \covmatbar{\mathbf Y}^{-1}  \crosscovmat{\mathbf X}{\mathbf Y} ^\dagger \right  ) \right ).   \label{eq-const-term-2}
\end{IEEEeqnarray} 
It remains to find   $\max I(S;\mathbf Y|\mathbf U)$ where the maximum is over all test channels $P_{S|\mathbf X}$ satisfying $I(S;\mathbf X|\mathbf U)\leq R$. \pe{ Let $\tilde{U} = U + \epsilon Z$ where $Z \sim (0, I)$. Applying the result of \cite[Theorem~2]{ZE19} on the triple $(X, Y, \tilde{U})$, which is Gaussian, and then taking the limit $\epsilon \to 0$ we get:}	
\begin{IEEEeqnarray}{rl}
\hspace{-1.6cm} \max_{\begin{subarray}{c} P_{S|\mathbf{X}} \: : \\ \hspace{0.9cm} I(S;\mathbf{X}|\mathbf{U}) \leq R  \end{subarray}} \!  \! \! \! \! \! \! \! \! \! \! \! \! \! \! &I(S;\mathbf{Y}|\mathbf{U}) = \max  \min \left \{ \!\! R + \frac{1}{2} \log  \big | \mathbf{I} - \mathbf \Omega \mathbf K_{\mathbf X | \mathbf Y} \big | \right. ; \nonumber \\
	 &\hspace{0cm} \frac{1}{2}\log \Big |  \mathbf{I}  +  \mathbf  \Omega  \crosscovmat{\mathbf X}{\mathbf Y}   \nonumber \\
	  &\hspace{0.2cm}  \left.  \times \Bigg (  \covmat{\mathbf Y}^{-1} -  \covmat{\mathbf Y}^{-1} \covmatbar{\mathbf Y}    \crosscovmatbar{\mathbf X}{\mathbf Y} ^{+} \crosscovmatbar{\mathbf X}{\mathbf Y} \covmatbar{\mathbf Y}^{-1}  \Bigg)   \crosscovmat{\mathbf X}{\mathbf Y} ^\dagger \Big |  \!\!\right \},\label{final-step}
\end{IEEEeqnarray} 
{ where the maximization in the last term is over all matrices $\mathbf 0 \preceq \mathbf  \Omega \preceq  \mathbf K_{\mathbf X | \mathbf Y}^{-1}$ and where $\bar{\mathbf K}_{\mathbf X \mathbf Y}^+$ designates the Moore-Penrose pseudo inverse of $\bar{\mathbf K}_{\mathbf X \mathbf Y}$.}

Summing \eqref{eq-const-term-1}--\eqref{final-step}  we obtain the desired result in \eqref{eq:result}, which completes the proof of Theorem~\ref{theorem-single-letter}.           

 \hfill\ensuremath{\blacksquare}
\section{Discussion }

In what follows, we show that constraint $\textnormal{C}$ as given by~\eqref{eq-condition-1} is \az{fulfilled} for a large class of sources even when $m=1$ and $q=2$. Let $X$ be a scalar source that is observed at the sensor and $\mathbf Y = (Y_1,Y_2)$ a $2$-dimensional source that is observed at the detector.
For convenience, let 
\begin{IEEEeqnarray}{r}
\mathbf{K} \!\!=\!\! \left [ \!\begin{matrix} \sigma^2_X & \sigma_{XY_1} & \sigma_{XY_2}  \\ \sigma_{XY_1} & \sigma^2_{Y_1} & \sigma_{Y_1Y_2}   \\  \sigma_{XY_2} & \sigma_{Y_1Y_2} & \sigma^2_{Y_2} \end{matrix} \! \right ] \textnormal{and }   \mathbf{\bar K}\!\!=\!\! \left  [ \! \begin{matrix} \bar \sigma^2_X & \bar \sigma_{XY_1} & \bar \sigma_{XY_2}  \\ \bar \sigma_{XY_1} & \bar \sigma^2_{Y_1} & \bar \sigma_{Y_1Y_2}  \\  \bar \sigma_{XY_2} & \bar \sigma_{Y_1Y_2} & \bar \sigma^2_{Y_2} \end{matrix} \! \right ]\!\!.\nonumber \\* 
\end{IEEEeqnarray}
Also, let 
\begin{equation}
a\!=\!\!\coefaa \textnormal{and }  b\!=\!\!\coefbb\!\!. \nonumber
\end{equation}
For this example the constraint $\textnormal{C}$ as given by~\eqref{eq-condition-1} reduces to 
\begin{subequations}\label{eq-equivalent-p-constrain}
	\begin{IEEEeqnarray}{rCl}
	i)\!\! &\quad& \sigma^2_X = \bar \sigma^2_X,  \\ 
	ii)\!\! &\quad&  \coefa (\sigma_{XY} - \bar \sigma_{XY})+ \coefb (\sigma_{XZ} - \bar \sigma_{XZ}) =0 \\
	iii)\!\! &\quad& \coefa^2(\sigma^2_{Y} - \bar \sigma^2_{Y} ) + 2 \coefa \coefb (\sigma_{YZ} - \bar \sigma_{YZ}) +\coefb^2 (\sigma^2_{Z} - \bar \sigma^2_{Z})= 0 \nonumber \\*
  \end{IEEEeqnarray}
\end{subequations}

For example, if all components have unit variance under both $P$ and $\bar P$, i.e., $\sigma_X^2 = \sigma_{Y_1}^2 = \sigma_{Y_2}^2 =1$ and $\bar \sigma_X^2 =\bar  \sigma_{Y_1}^2 =\bar  \sigma_{Y_2}^2 =1$ then all definite positive matrices $\mathbf K$ and $\bar{ \mathbf K}$ of the form 
\begin{equation}
 \mathbf{K}= \left [ \begin{matrix}1&  a_{12} &   { \color{black} h ( \bar a_{12} , \bar a_{13} , \bar a_{23}, a_{12}) } \\  a_{12} & 1 &  { \color{black} \bar a_{23} } \\ { \color{black} h ( \bar a_{12} , \bar a_{13} ,   \bar a_{23} , a_{12}) }   &   { \color{black} \bar a_{23} } & 1 \end{matrix}\right ],\end{equation}
 and 
 \begin{equation} \mathbf{\bar K} = \left [ \begin{matrix}1&   a_{12} & \bar a_{13} \\ a_{12} & 1 & \bar a_{23} \\ \bar a_{13}  &  \bar a_{23} & 1 \end{matrix}\right ]
\end{equation}
for some arbitrary parameters $a_{12}$, $\bar a_{12}$, $\bar a_{13}$, $\bar a_{23}$, satisfy the constraint~\eqref{eq-equivalent-p-constrain}. Here 
\begin{equation}
h(x,y_1,y_2,t) = y_1 - (t-y_2)\frac{y_1y_2 - x}{ xy_2 -y_1}.
\end{equation}
\begin{example}\label{example-p2P-optimal}
	Let 
	\mw{
	\begin{equation}
	\mathbf{  K}= \left [ \begin{matrix}1&  0.4 & \alpha \\  0.4 & 1 & 0.1 \\ \alpha  &  0.1 & 1  \end{matrix}\right ] \quad \textnormal{and}\quad  \mathbf{\bar K} =\left [ \begin{matrix}1&  0.1 &-0.8 \\  0.1 & 1 & 0.1 \\ -0.8  &  0.1 & 1  \end{matrix}\right ] ,
	\end{equation}}
	with $\alpha \approx -0.73333$.
	It is easy to see that~\eqref{eq-equivalent-p-constrain} is fulfilled. Figure~\ref{fig-optimum-gaussian} shows the evolution of the optimal exponent $E$ as a function of the communication rate R as given by Theorem~\ref{theorem-single-letter} for this example. \mw{Notice  that Han's exponent \cite[Theorem 2]{H87} is strictly suboptimal for this example\footnote{In the figure, Han's exponent as given by \cite[Theorem 2]{H87} is computed using Gaussian test channels $P_{U|X}$ and Gaussian $\tilde{U}$}.}
	\begin{figure}[ht]
	\begin{center}
		\scalebox{0.5}{\input{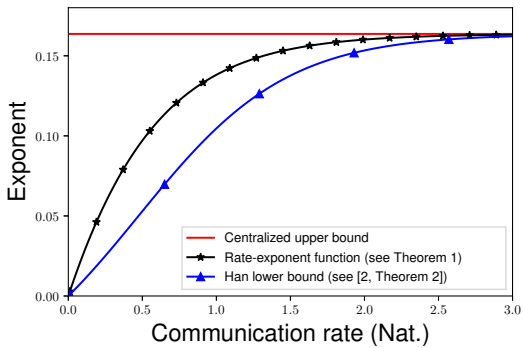}}
	\end{center}
\vspace{-.5cm}

	\caption{Rate-exponent region for Example~\ref{example-p2P-optimal}.}
	\label{fig-optimum-gaussian}
\end{figure}
\end{example}
\section*{Acknowledgement}
The work of M. Wigger was funded by the European Research Council (ERC) under the European Union's Horizon 2020 under grant agreement No 715111.
\bibliographystyle{IEEEtran}
\bibliography{IEEEabrv,./root}

\begin{thebibliography}{10}
\providecommand{\url}[1]{#1}
\csname url@samestyle\endcsname
\providecommand{\newblock}{\relax}
\providecommand{\bibinfo}[2]{#2}
\providecommand{\BIBentrySTDinterwordspacing}{\spaceskip=0pt\relax}
\providecommand{\BIBentryALTinterwordstretchfactor}{4}
\providecommand{\BIBentryALTinterwordspacing}{\spaceskip=\fontdimen2\font plus
\BIBentryALTinterwordstretchfactor\fontdimen3\font minus
  \fontdimen4\font\relax}
\providecommand{\BIBforeignlanguage}[2]{{%
\expandafter\ifx\csname l@#1\endcsname\relax
\typeout{** WARNING: IEEEtran.bst: No hyphenation pattern has been}%
\typeout{** loaded for the language `#1'. Using the pattern for}%
\typeout{** the default language instead.}%
\else
\language=\csname l@#1\endcsname
\fi
#2}}
\providecommand{\BIBdecl}{\relax}
\BIBdecl

\bibitem{AC86}
R.~Ahlswede and I.~Csiszar, ``Hypothesis testing with communication
  constraints,'' \emph{IEEE Trans. Inf. Theory}, vol.~32, no.~4, pp. 533--542,
  July 1986.

\bibitem{H87}
T.~S. Han, ``Hypothesis testing with multiterminal data compression,''
  \emph{IEEE Trans. Inf. Theory}, vol.~33, no.~6, pp. 759--772, November 1987.

\bibitem{SHA94}
H.~Shimokawa, T.~S. Han, and S.~Amari, ``Error bound of hypothesis testing with
  data compression,'' in \emph{Proc. {IEEE} {ISIT'94}}, Jun. 1994, p. 114.

\bibitem{RW12}
M.~S. Rahman and A.~B. Wagner, ``On the optimality of binning for distributed
  hypothesis testing,'' \emph{IEEE Trans. Inf. Theory}, vol.~58, no.~10, pp.
  6282--6303, Oct. 2012.

\bibitem{ZE19}
A.~Zaidi and I.~E. Aguerri, ``Optimal rate-exponent region for a class of
  hypothesis testing against conditional independence problems,'' in \emph{2019
  IEEE Information Theory Workshop (ITW)}, Aug 2019, pp. 1--5.

\bibitem{ZL14A}
W.~{Zhao} and L.~{Lai}, ``Distributed testing against independence with
  multiple terminals,'' in \emph{2014 52nd Annual Allerton Conference on
  Communication, Control, and Computing (Allerton)}, Sep. 2014, pp. 1246--1251.

\bibitem{SWT18}
S.~{Salehkalaibar}, M.~{Wigger}, and R.~{Timo}, ``On hypothesis testing against
  conditional independence with multiple decision centers,'' \emph{IEEE
  Transactions on Communications}, vol.~66, no.~6, pp. 2409--2420, June 2018.

\bibitem{EWZ19}
P.~Escamilla, M.~Wigger, and A.~Zaidi, ``Distributed hypothesis testing:
  cooperation and concurrent detection,'' \emph{in revision for publication in
  the IEEE Transactions of Information Theory}, 2019.

\bibitem{TC08}
C.~{Tian} and J.~{Chen}, ``Successive refinement for hypothesis testing and
  lossless one-helper problem,'' \emph{IEEE Transactions on Information
  Theory}, vol.~54, no.~10, pp. 4666--4681, Oct 2008.

\bibitem{XK12}
Y.~Xiang and Y.~H. Kim, ``Interactive hypothesis testing with communication
  constraints,'' in \emph{Proc. of \emph{Allerton Conference on Comm., Control,
  and Comp.}}, Monticello (IL), USA, Oct. 2012, pp. 1065--1072.

\bibitem{KPD16A}
G.~{Katz}, P.~{Piantanida}, and M.~{Debbah}, ``Collaborative distributed
  hypothesis testing with general hypotheses,'' in \emph{2016 IEEE
  International Symposium on Information Theory (ISIT)}, July 2016, pp.
  1705--1709.

\bibitem{WT16}
M.~Wigger and R.~Timo, ``Testing against independence with multiple decision
  centers,'' in \emph{2016 International Conference on Signal Processing and
  Communications (SPCOM)}, Bangalore, India, Jun. 2016, pp. 1--5.

\bibitem{SWW17}
S.~Salehkalaibar, M.~A. Wigger, and L.~Wang, ``Hypothesis testing in multi-hop
  networks,'' \emph{arXiv:1708.05198}, 2017.

\bibitem{EWZ18}
P.~{Escamilla}, M.~{Wigger}, and A.~{Zaidi}, ``Distributed hypothesis testing
  with concurrent detections,'' in \emph{2018 IEEE International Symposium on
  Information Theory (ISIT)}, June 2018, pp. 166--170.

\bibitem{ZL18}
W.~{Zhao} and L.~{Lai}, ``Distributed testing with cascaded encoders,''
  \emph{IEEE Transactions on Information Theory}, vol.~64, no.~11, pp.
  7339--7348, Nov 2018.

\bibitem{EZW19}
P.~Escamilla, A.~{Zaidi}, and M.~{Wigger}, ``Distributed hypothesis testing
  with collaborative detection,'' in \emph{2018 56th Annual Allerton Conference
  on Communication, Control, and Computing (Allerton)}, Oct 2018, pp. 512--518.

\bibitem{SW18A}
S.~{Salehkalaibar} and M.~{Wigger}, ``Distributed hypothesis testing over
  multi-access channels,'' in \emph{2018 Information Theory and Applications
  Workshop (ITA)}, Feb 2018, pp. 1--5.

\bibitem{SG17}
S.~{Sreekumar} and D.~{G\"und\"uz}, ``Distributed hypothesis testing over noisy
  channels,'' in \emph{2017 IEEE International Symposium on Information Theory
  (ISIT)}, June 2017, pp. 983--987.

\bibitem{LSCV17}
J.~Liao, L.~Sankar, F.~P. Calmon, and V.~Y. Tan, ``Hypothesis testing under
  maximal leakage privacy constraints,'' in \emph{2017 IEEE International
  Symposium on Information Theory (ISIT)}.\hskip 1em plus 0.5em minus
  0.4em\relax IEEE, 2017, pp. 779--783.

\bibitem{LSTC18}
J.~{Liao}, L.~{Sankar}, V.~Y.~F. {Tan}, and F.~{du Pin Calmon}, ``Hypothesis
  testing under mutual information privacy constraints in the high privacy
  regime,'' \emph{IEEE Transactions on Information Forensics and Security},
  vol.~13, no.~4, pp. 1058--1071, April 2018.

\bibitem{SGC18}
S.~{Sreekumar}, D.~{G\"und\"uz}, and A.~{Cohen}, ``Distributed hypothesis
  testing under privacy constraints,'' in \emph{2018 IEEE Information Theory
  Workshop (ITW)}, Nov 2018, pp. 1--5.

\bibitem{SWW19}
S.~{Salehkalaibar}, M.~{Wigger}, and L.~{Wang}, ``Hypothesis testing over the
  two-hop relay network,'' \emph{IEEE Transactions on Information Theory},
  vol.~65, no.~7, pp. 4411--4433, July 2019.

\end{thebibliography}
\end{document}